\documentclass[a4paper, aps, prl, longbibliography, reprint, nofootinbib]{revtex4-1}

\usepackage{asymptote} 
\usepackage{array}
\usepackage{amsmath,amssymb,amsfonts}
\usepackage[colorlinks]{hyperref}
\usepackage{multirow}
\usepackage{float} 
\usepackage[utf8]{inputenc}
\usepackage{graphicx}
\usepackage{subfigure}
\graphicspath{{./pic/}}

\usepackage[utf8]{inputenc}
\newcommand{\bw}{\begin{widetext}}
\newcommand{\ew}{\end{widetext}}
\newcommand{\be}{\begin{equation}}
\newcommand{\en}{\end{equation}}
\newcommand{\bee}{\begin{equation}}
\newcommand{\ene}{\end{equation}}
\newcommand{\bea}{\begin{eqnarray}}
\newcommand{\ena}{\end{eqnarray}}
\newcommand{\bes}{\begin{subequations}}
\newcommand{\ens}{\end{subequations}}
\newcommand{\bef}{\begin{figure}}
\newcommand{\enf}{\end{figure}}
\newcommand{\eq}[1]{Eq.~(\ref{#1})}

\def\ms{\;\;\,}

\def\ie{{\it i.e.}}

\def\cala{\mathcal{A}}

\def\calc{\mathcal{C}}
\def\cald{\mathcal{D}}

\def\calg{\mathcal{G}}

\def\call{\mathcal{L}}
\def\calm{\mathcal{M}}

\def\calp{\mathcal{P}}

\def\to{\rightarrow}

\def\d{{\rm d}}
\def\tev{{\rm TeV}}
\def\gev{{\rm GeV}}

\begin{document}


\title{
	Enhancing $CP$ Measurement of the Yukawa Interactions of Top-Quark at $e^{-}e^{+}$  Collider
}

\author{Kai Ma}
\email[Electronic address: ]{makainca@yeah.net}
\affiliation{School of Physics Science, Shaanxi University of Technology, Hanzhong 723000, Shaanxi, China}

\date{\today}

\begin{abstract}
It is well known that $CP$-violating coupling of the Yukawa interaction of top-quark is a promising candidate of a new source of $CP$ violation effect. Precisely measurement of its $CP$ properties is crucial for understanding new physics above the electroweak scale. In this paper, we introduce a complete analysis method for probing $CP$ violation effects in the associated production of top-quark pair and Higgs boson at $e^{-}e^{+}$ collider. Reconstructions of the top-quarks are not needed in our strategy. The observables are defined based on spin correlation effects of the leptons emerging from decays of top-quarks, and their formal expressions are universal at any reference frames. Two reference frames, the rest frame of the top-quark pair and the rest frame of Higgs boson, are examined. Importantly, large enhancement effects for both $CP$-odd and $CP$-even observables are observed in the rest frame of Higgs boson. This can be essential for probing $CP$ violation effects at future $e^{-}e^{+}$ collider.

\end{abstract}


\maketitle


%
After the long waited era in the elementary particle physics, a scalar particle with mass $\sim125\gev$~\cite{ATLAS:CMS:2015} was discovered at LHC in 2012~\cite{Aad:2012tfa,Chatrchyan:2012xdj}. However, our knowledge about its physical properties can not be less any more to make a judgment that it is really the Higgs boson who spontaneously breaks the electroweak vacuum and gives masses to the observed vector bosons and fermions, because so far its self-coupling which is responsible for the symmetry breaking has not been measured~\cite{TheATLASandCMSCollaborations:2015bln}. On the other hand, theoretical extensions of the Standard Model (SM) always possess more than one Higgs doublet, and hence more scalar particles with either heavier or lighter masses, but having $CP$ quantum numbers different from the SM prediction, $J^{CP}=0^{++}$, are predicted. Therefore, the observed mass eigenstate $h(125)$ (we will use this notation in the rest of this paper) can be potentially a mixture of the $CP$ eigenstates. As a natural expectation that the Higgs boson predicted by the SM is at least the dominant component of the observed scalar particle $h(125)$, decisive measurement on the $CP$ violation effect, which is a promising potential source of the observed  asymmetry between matter and anti-matter in our universe~\cite{Sakharov:1967dj, Espinosa:2011eu, Shu:2013uua}, is important and urgent for understanding the physics above electroweak scale. However, it is challenging in practice~\cite{Hagiwara:2016zqz}.

Without loss of generality, we can assume that $h(125)$ is a superposition of $CP$-even eigenstate $H$ and $CP$-odd eigenstate $A$, 
\bee\label{eq:mixing}
h = H\cos\xi  + A\sin\xi  \,,
\ene
where $\xi$ is the Higgs mixing angle that has been assumed to be real. 
The story is further complicated by the couplings of $H$ and $A$ to the observed vector bosons and fermions because what we measure are usually combinations of those couplings and the mixing angle $\xi$. Therefore,  in general, channel by channel studies on the $CP$ violation effects related to $h(125)$ is unavoidable. 

Soon after $h(125)$ was observed, its $CP$ property was studied through $h\to ZZ\to4\ell$ by the ATLAS and CMS collaborations \cite{Aad:2013xqa,Chatrchyan:2012jja,Chatrchyan:2013mxa}, and the results disfavor the $CP$-odd hypothesis by nearly $3\sigma$, which is a rather weak constraint on the mixing angle due to the complexity mentioned above. On the other hand, while $HVV'$ couplings usually appear at tree level, the $AVV'$ interactions are only loop induced, and hence $CP$ violation is heavily suppressed~\cite{Plehn:2001nj, Hagiwara:2009wt,Ellis:2012jv,Bernlochner:2018opw}. This drawback absents, theoretically, in the couplings of $H$ and $A$ to the fermions, where both types of interaction can happen at tree level and hence are promising probes of $CP$ violation effects~\cite{Ellis:2013yxa}. Recently, The decay of $h(125)$ to $\tau$-lepton pair channel was analyzed by the ATLAS and CMS collaborations~\cite{Aad:2016nal, Zanzi:2017msx}, and the bound is again rather weak. The optimization method proposed in Ref.~\cite{Hagiwara:2016zqz} is expected to enhance the sensitivity to a promising level. However, the precision is still limited for inferring physics beyond SM. It is expected that the sensitivity can be significantly improved at future $e^{-}e^{+}$ collider~\cite{Li:2016zzh}. However, since $h(125)$ are produced (at relatively low energy) through the strahlung process $e^{-}e^{+}\to h Z$, its $CP$-odd component can be completely wash out due to the inherent smallness of the $AZZ$ coupling~\cite{BhupalDev:2007ftb,Godbole:2007cn}. The $CP$ quantum numbers can be determined unambiguously at $\gamma\gamma$ collider~\cite{Grzadkowski:1992sa, Gunion:1994wy}, but this is not a practical option since it is too remote. 

Many efforts have also been devoted to study possible $CP$ violation effects in the couplings of $h(125)$ to top-quark at the LHC~\cite{Gunion:1996vv, Demartin:2014fia,  Kobakhidze:2014gqa, Kobakhidze:2016mfx, Li:2017dyz, Rindani:2016scj, Mileo:2016mxg, Yue:2014tya, Bernreuther:1993df, Buckley:2015vsa,Godbole:2011hw,Boudjema:2015nda,Goncalves:2018agy,Barger:2018tqn}. Higgs boson production in association with a top-quark pair has been observed recently by both CMS and ATLAS collaborations~\cite{Sirunyan:2018hoz, Aaboud:2018urx}, 
however the expected sensitivity to $CP$-violating parameter is not much powerful compared to the constraints from electric dipole moments~\cite{Brod:2013cka,Chen:2015gaa}. The analysis in Ref.~\cite{Goncalves:2018agy} shown that at the high luminosity LHC ($3{\rm ab}^{-1}$) the SM Higgs can be distinguished at 95\% confidence level from a $CP$-mixed state with $\cos\xi_{ht\bar{t}} < 0.7$, and observation of the sign of the mixing parameter with maximum mixing requires at least $8{\rm ab}^{-1}$ of data at $1\sigma$-level. Similar results were also reported in Ref.~\cite{Buckley:2015vsa}. Even through improvements may be possible, it is still challenging to study the details of the $CP$-violating parameter. The $CP$ violating interactions can also be investigated via loop-induced process. The $pp \to h+2j$ process, which are second order corrections to the gluon fusion process, is of particularly interesting~\cite{Englert:2019xhk}. The events are generated dominantly via the top-quark running in the loop~\cite{Kauffman:1996ix,Kauffman:1998yg}, and hence $CP$ violating coupling can affect both total cross section and angular distributions of the jets~\cite{Dolan:2014upa,DelDuca:2006hk, Klamke:2007cu}. The analysis in Ref.~\cite{Englert:2019xhk} shown that compared to the $ht\bar{t}$ process, the $h+2j$ events can give much strong constraint on the $CP$ violating parameter. However, the bound depends heavily on the (effective) couplings between the Higgs and gluons.


In contrast, the $CP$-even and $CP$-odd components are generated fairly in the associated production of $h(125)$ and top-quark pair, $e^{-}e^{+}\to ht\bar{t}$. On the other hand, many extensions of the SM predict that the heavier the particle, the stronger the interactions with the unknown sector. The top-quark being the heaviest particle that has been observed, is hence a natural probe of new physics above the electroweak scale~\cite{Fedor:2017,Markus:2017,Ulrich:2017,Cirigliano:2016njn}. Therefore, most likely possible $CP$ violation effects in the Higgs sector will be demonstrated in first in the interactions between $h(125)$ and top-quark~\cite{Gunion:1996vv, Ananthanarayan:2014eea, BhupalDev:2007ftb, Li:2016zzh,Bernreuther:2017cyi}. 

The associated production process $e^{-}e^{+}\to ht\bar{t}$ is dominantly generated by radiation of the Higgs boson from one of the top-quarks. Contribution of the production in association with a $Z$ boson, $e^{-}e^{+}\to hZ \to ht\bar{t}$, is very small, only a few percent when $\sqrt{s} \le 1\tev$~\cite{Hagiwara:2016rdv}. Meanwhile, the photon exchange channel contributes the bulk of the cross section. It has been shown that the total cross section and the top-quark polarization would be good observables for measuring $CP$ violation parameter~\cite{BhupalDev:2007ftb}. On the other hand, Since the top-quark decays before hadronization, its spin-polarization can be measured by studying the polar angle~\cite{Jezabek:1989} as well as the azimuthal angle distributions~\cite{Godbole:2010kr} of its decay products, particularly the lepton. Furthermore, the associated spin correlations between the top-quark and anti-top-quark provide much richer physics for probing anomalous interactions in the production dynamics~\cite{Aguilar-Saavedra:2014kpa}. However, previously works studied either $CP$-even observables~\cite{BhupalDev:2007ftb, Ellis:2013yxa} which is insensitive to the $CP$ violation parameter, or $CP$-odd observables which requiring reconstruction of the momenta of top-quark and anti-top-quark~\cite{Gunion:1996vv, Antipin:2008zx, Ananthanarayan:2014eea, BhupalDev:2007ftb, Li:2016zzh, Hagiwara:2017ban}.


Recently, a general method for analyzing possible non-trivial spin correlations by using angular correlations between leptons emerging from the top-quark pair decay was proposed~\cite{Ma:2017vve}. Based on these distributions, a complete set of asymmetry observables is constructed for investigating new physics in the production dynamics. It has been shown that the angular distributions of the leptons from top-quark and anti-top-quark decays are insensitive to new physics in the decay side~\cite{Godbole:2006tq}, and furthermore non-trivial dependence of the energy distributions are completely removed in the asymmetry observables due to its definition. In consideration of these advantages, in this paper, we propose a complete set of asymmetry observables to study the $CP$ violation effects in the process $e^{-}e^{+}\to ht\bar{t}$.

For our discussion, we assume that Yukawa interactions between the $CP$-even eigenstate $H$ and the $CP$-odd eigenstate $A$ and the top-quark are $CP$ conserving, such that the Higgs mixing parameter $\xi$ in \eq{eq:mixing} is the only source of $CP$ violation in the Higgs sector. In this case, the corresponding  Lagrangian can be parameterized as,
\bee
\call_{{\rm int.}} 
= - g_{Ht\bar{t}}\;\overline{\psi}_{t}\psi_{t} \, H -  i g_{At\bar{t}}\;\overline{\psi}_{t}\gamma^{5}\psi_{t}\, A \,.
\ene
For the associated production process $e^{-}e^{+}\to ht\bar{t}$, production rates of the $CP$-even and $CP$-odd components in the mass eigenstate $h(125)$ are subjected to the above Lagrangian and the mixing parameter $\xi$. Since $h(125)$ will be looked into inclusively, we don't need to consider the projection effect~\cite{BhupalDev:2007ftb} in its decay (in practice, usually we can not investigate all decay channels, in this case projection effect has to be taken into account for precision measurement). Therefore, the interactions between the mass eigenstate $h(125)$ and the top-quark can be described as,
\bee
\call_{\rm int.} 
=
- g_{Ht\bar{t}}\; h \big( \cos\xi\, \overline{\psi}_{t}\psi_{t} + i \kappa_{ht\bar{t}}\sin\xi\,\overline{\psi}_{t}\gamma^{5}\psi_{t} \big) \,,
\ene 
where $\kappa_{ht\bar{t}} = g_{At\bar{t}} / g_{Ht\bar{t}} $. In general $\kappa_{ht\bar{t}}$ can be arbitrary in both its magnitude and sign, and hence model-dependent. The strength of $CP$ violation effect can also be affected by kinematic factors. For instance, the $CP$ violating phase measured through the observables proposed in Ref.~\cite{Hagiwara:2016rdv} is suppressed by a factor of $\sim 0.2$. In this paper, a model-independent assumption $\kappa_{ht\bar{t}}=+1$ will be continuously-used unless it is mentioned. Furthermore, we will take $g_{Ht\bar{t}}$ as the Yukawa coupling of top-quark in the SM. 

We will study the spin correlation effects in the leptonic decays of top-quarks. For the process $e^{-}e^{+}\to ht\bar{t}$, the differential cross section can be written as~\cite{Ma:2017vve},
\bee
\frac{ \d\sigma }{ \d\cos\theta_{1}\d\phi_{1}\d\cos\theta_{2}\d\phi_{2} }  
= \sum_{ \lambda_{f}, \lambda_{f}' }
\calp^{ \lambda_{f} }_{ \lambda_{f}' } \cald^{ \lambda_{f} }_{ \lambda_{f}' }(\theta_{i}, \phi_{i} ) \,,
\ene
where $\theta_{i}$ and $\phi_{i}$ with $i=1$ ($i=2$) are the polar angle and azimuthal angle of anti-lepton (lepton) in a chosen reference frame $R$; $\lambda_{f}=s, 0, \pm1$ are the helicity of the lepton-pair system projected along the $z$-axis of the reference frame $R$; $\calp^{ \lambda_{f} }_{ \lambda_{f}' }$ and $\cald^{ \lambda_{f} }_{ \lambda_{f}' }(\theta_{i}, \phi_{i} )$ stand for the spin-projected production density matrix and the correlations functions of the leptonic pair system. Within the SM, the correlation function $\cald^{ \lambda_{f} }_{ \lambda_{f}' }(\theta_{i}, \phi_{i} )$ is universal for any production dynamics, and has been given in Ref.~\cite{Ma:2017vve}. On the other hand, the spin-projected production helicity amplitudes can be generally written as
\bee\label{eq:hel:production}
\calm_{P}( \lambda, \lambda_{f} ) = \sum_{\lambda_{f}}  \calg^{\mu\nu}(\lambda_{f}) L_{\mu}(\lambda) \epsilon_{\nu}(\lambda_{f})\,,
\ene
where $L_{\mu}(\lambda)$ is current of the initial electron pair, and $\lambda$ stands for its polarization (since beam polarization effect is small for the most sensitive observables~\cite{Bernreuther:1995nw}, we will don't discuss this part in this paper.); $\epsilon_{\nu}(\lambda_{f})$ is the spin-projection factor, and $\calg(\lambda_{f})^{\mu\nu}$ accounts for Lorentz structures determined by the production dynamics. 

Apart from the total cross section, which can be good observable before one reaches the chiral limit at very high energies, 15 additional asymmetry observables can be defined based on the correlation function $\cald^{ \lambda_{f} }_{ \lambda_{f}' }(\theta_{i}, \phi_{i} )$. The magnitudes of these asymmetry observables are determined by the spin-projected density matrix $\calp^{ \lambda_{f} }_{ \lambda_{f}' }$, the analytical results are too lengthy and will be given elsewhere~\cite{Ma:2018www}.
In terms of the number of events in corresponding phase space regions, the asymmetry observables can defined as follows,
\bes
\bea
\calc_{\phi_{i}}
&=&
\frac{ N(\sin\phi_{i} > 0) - N(\sin\phi_{i} < 0)}{ N(\sin\phi_{i} > 0) + N(\sin\phi_{i} < 0) }\,,
\\[2mm]
\calc_{\phi_{\pm}}
&=&
\frac{ N(\sin2\phi_{\pm} > 0) - N(\sin2\phi_{\pm} < 0)}{ N(\sin2\phi_{\pm} > 0) + N(\sin2\phi_{\pm} < 0) }\,,
\\[2mm]
\calc_{ \theta_{i}\phi_{i'} }
&=&
\frac{ N( \cos\theta_{i}\sin\phi_{i'} > 0) - N( \cos\theta_{i}\sin\phi_{i'} < 0)}{ N( \cos\theta_{i}\sin\phi_{i'} > 0) + N( \cos\theta_{i}\sin\phi_{i'} < 0) }\,,
\ena
\bea
\cala_{\theta_{i} }
&=&
\frac{ N(\cos\theta_{i} > 0) - N(\cos\theta_{i} < 0)}{ N(\cos\theta_{i} > 0) + N(\cos\theta_{i} < 0) }\,,
\\[2mm]
\cala_{\phi_{i}}
&=&
\frac{ N(\cos\phi_{i} > 0) - N(\cos\phi_{i} < 0)}{ N(\cos\phi_{i} > 0) + N(\cos\phi_{i} < 0) }\,,
\\[2mm]
\cala_{\theta_{1}\theta_{2}}
&=&
\frac{ N(\cos\theta_{1}\cos\theta_{2} > 0) - N(\cos\theta_{1}\cos\theta_{2} < 0)}{ N(\cos\theta_{1}\cos\theta_{2} > 0) + N(\cos\theta_{1}\cos\theta_{2} < 0) }\,,
\\[2mm]
\cala_{\phi_{\pm}}
&=&
\frac{ N(\cos2\phi_{\pm} > 0) - N(\cos2\phi_{\pm} < 0)}{ N(\cos2\phi_{\pm} > 0) + N(\cos2\phi_{\pm} < 0) }\,,
\\[2mm]
\cala_{\theta_{i}\phi_{i'} }
&=&
\frac{ N( \cos\theta_{i}\cos\phi_{i'} > 0) - N( \cos\theta_{i}\cos\phi_{i'} < 0)}{ N( \cos\theta_{i}\cos\phi_{i'} > 0) + N( \cos\theta_{i}\cos\phi_{i'} < 0) }\,,
\ena
\ens
where $\phi_{\pm}= (\phi_{1}\pm\phi_{2})/2$. The first three equations give 6 $CP$-odd observables, and the last five equations give 9 $CP$-even observables.

Most importantly, the asymmetry observables listed above are valid in any reference frame~\cite{Ma:2017vve}. Because it is the left-handed current through which top-quark and anti-top-quark decay, the spin-projected production density matrix elements and hence the asymmetry observables have a relatively strong dependence on the reference frame. Therefore, performing optimization for finding the most sensitive asymmetry observables in various reference frames, which is also one of the advantages of our approach, is important. In this paper, we consider two reference frames that can be determined directly without of reconstructions of top-quarks: 1) the rest frame of the top-quark pair system, denoted by $R_{\psi}$; 2) the rest frame of the Higgs boson, denoted by $R_{h}$. Definitions of the axes of these two reference frames are given in the caption of Fig.~\ref{fig:frame}.
\begin{widetext} 
\mbox{}
\begin{figure}[t]
	\begin{center}
		\subfigure[]{\label{fig:frame:psi}}
		{\includegraphics[scale=0.80]{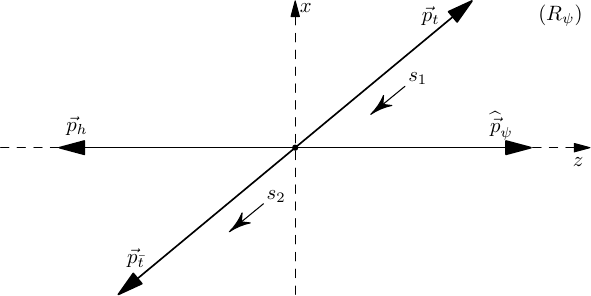}}	
		\hfill
		\subfigure[]{\label{fig:frame:hig}}
		{\includegraphics[scale=0.80]{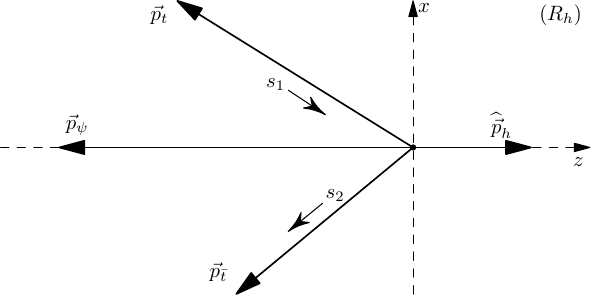}}
	\end{center}
	\caption{Dominant geometry configurations of the momenta and helicity of the top-quark pair in the rest frames of (a) the top-quark pair and (b) the scalar particle $h(125)$. The $z$ axis of $R_{\psi}$ is along the flying direction of the top-quark pair system, and the $x$ axis is determined by the plane spanned by the momentum of $e^{-}$ and the $z$ axis. The $z$ axis of $R_{h}$ is along the flying direction of $h(125)$, and the $x$ axis is determined by the plane spanned by the momentum of $e^{-}$ and he $z$ axis.}
	\label{fig:frame}
\end{figure}
\end{widetext}
\begin{figure}[t]
	\begin{center}
	        \subfigure[]{\label{fig:xs:oddPhi}}
		{\includegraphics[scale=0.56]{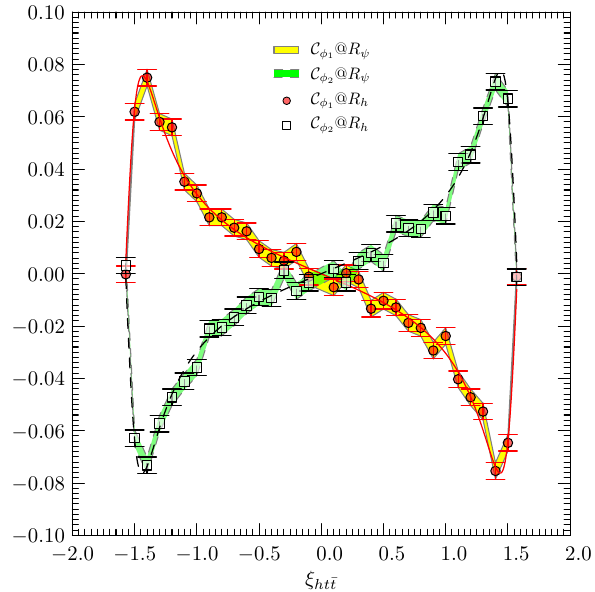}}	
		\subfigure[]{\label{fig:xs:oddPhiPhi}}
		{\includegraphics[scale=0.56]{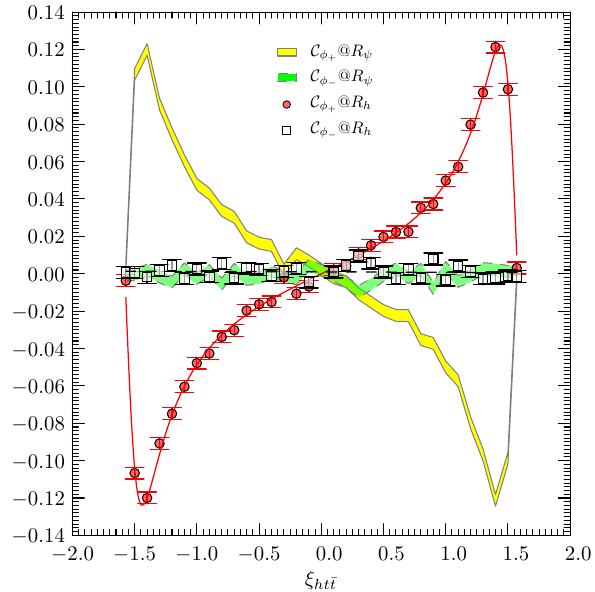}}
		\subfigure[]{\label{fig:xs:oddThePhi}}
		{\includegraphics[scale=0.56]{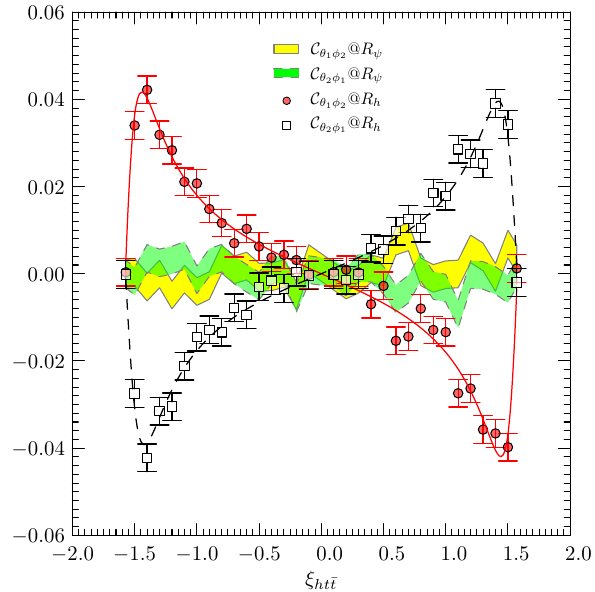}}
	\caption{Dependence of the $CP$-even asymmetry observables on the $CP$ violation parameter for $\sqrt{s}=500\gev$. For illustration, $10^{5}$ events are generated by using MadGraph~\cite{Alwall:2014hca}. 
	}	
	\label{fig:xs-C}
	\end{center}
\end{figure}

Fig.~\ref{fig:xs:oddPhi}, \ref{fig:xs:oddPhiPhi} and \ref{fig:xs:oddThePhi} show the dependences of the above 6 $CP$-odd asymmetry observables $\calc_{\phi_{i}}$, $\calc_{\phi_{\pm}}$ and $\calc_{ \theta_{i}\phi_{i'} }$. Among them, $\calc_{\phi_{i}}$ and $\calc_{\phi_{+}}$ defined in the reference frame $R_{\psi}$ have been predicted in the Ref.~\cite{Hagiwara:2016rdv}. Because they are transverse correlations among azimuthal angles of the leptons, boost along the $z$-axis does not affect their values as expected. However, the observables $\calc_{ \theta_{i}\phi_{i'} }$ strongly depend on the boost due to that they are related to the polar angle. This can been clearly seen in Fig.~\ref{fig:xs:oddThePhi}: in the reference frame $R_{h}$, $\calc_{ \theta_{i}\phi_{i'} }$ are sensitive to the $CP$ violation parameter $\xi$, but they are trivial in the reference frame $R_{\psi}$.
\begin{widetext}
\mbox{}
\begin{figure}[t]
	\begin{center}
		\subfigure[]{\label{fig:xs:evenThe}}
		{\includegraphics[scale=0.56]{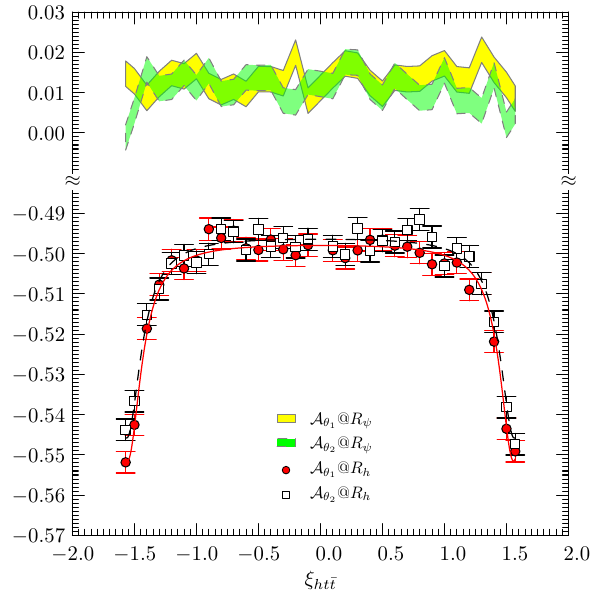}}
		\subfigure[]{\label{fig:xs:evenPhi}}
		{\includegraphics[scale=0.56]{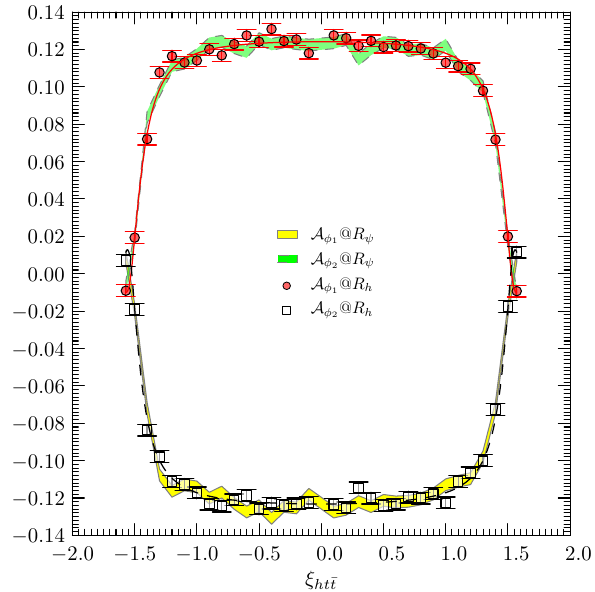}}
		\subfigure[]{\label{fig:xs:evenTheThe}}
		{\includegraphics[scale=0.56]{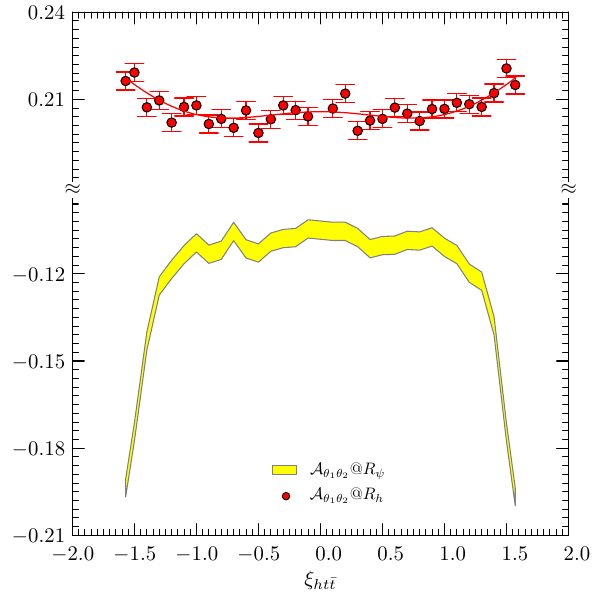}}
		\\
		\subfigure[]{\label{fig:xs:evenPhiPhi}}
		{\includegraphics[scale=0.56]{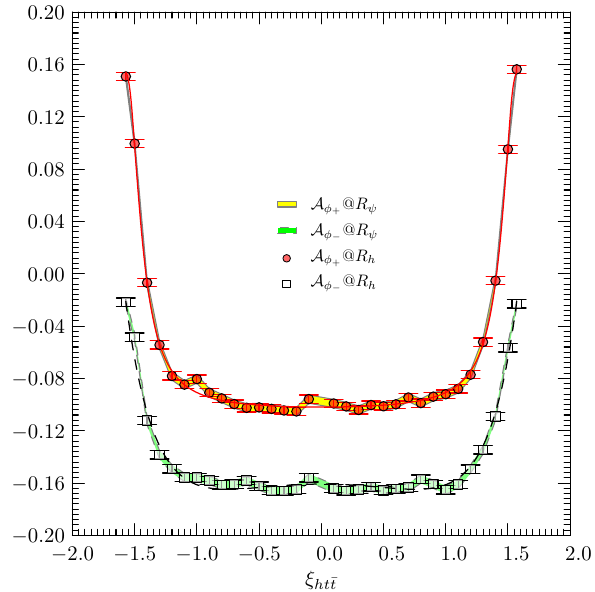}}
		\subfigure[]{\label{fig:xs:evenThePhi}}
		{\includegraphics[scale=0.56]{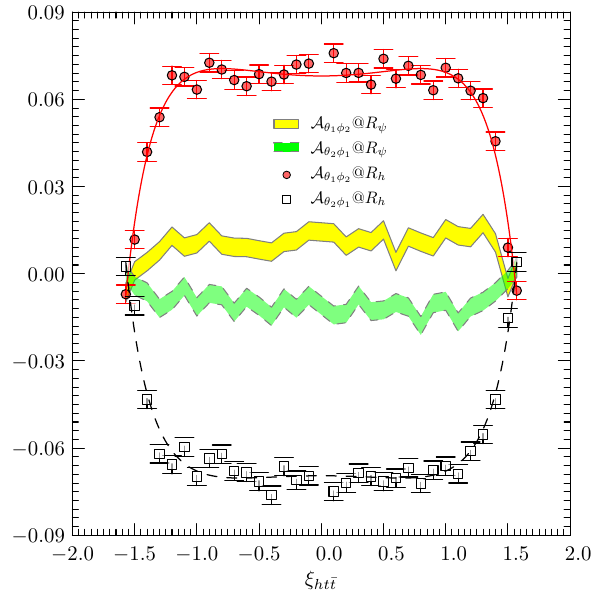}}
        \vspace{-0.3cm}
	\caption{Dependence of the $CP$-even asymmetry observables on the $CP$ violation parameter for $\sqrt{s}=500\gev$.}	
	\label{fig:xs}
	\end{center}
\end{figure}
\end{widetext}

This enhancement effect can be understood by the different kinematical configurations in these two reference frames, see Fig.~\ref{fig:frame}. Firstly, the observables $\calc_{ \theta_{i}\phi_{i'} }$ are due to interferences between the transverse and longitudinal as well as scalar spin-projected components~\cite{Ma:2017vve}, \ie, proportional to $|\calp^{\pm}_{0}|=|\calm_{P}^{\pm}\calm_{P}^{0}|$ and $|\calp^{\pm}_{s}|=|\calm_{P}^{\pm}\calm_{P}^{s}|$.
In the reference frame $R_{\psi}$, where the top-quark and anti-top-quark are back-to-back, because the top-quarks are left-handed, the transversely spin-projected density matrix elements $|\calp^{-}_{-}|$ and $|\calp^{+}_{+}|$ are dominant, while the longitudinal- and singlet-projected density matrix elements, $|\calp_{0}^{0}|$ and $|\calp_{s}^{s}|$, are relatively small. In contrast, in the reference frame $R_{h}$, the top-quark pair system is boosted along the negative $z$ direction. Therefore, most likely, both the top-quark and anti-top-quark are in the negative side of the $z$-axis, and hence, the longitudinal and singlet spin-projected density matrix elements, $|\calp_{0}^{0}|$ and $|\calp_{s}^{s}|$ are relatively larger. The above property can also been seen via the eq.~(\ref{eq:hel:production}). For instance, in case of that $\calg_{\mu\nu} \propto g_{\mu\nu}$, then the production helicity amplitudes with $\lambda_{f}=s$ in these two reference frames are given as,
\bee
\calm_{P}(\lambda_{f}=s) \propto 
\begin{cases}
\gamma_{\psi}\beta_{\psi} & \text{in $R_{\psi}$}\,,
\\[2mm]
\gamma_{h}\beta_{h} &  \text{in $R_{h}$}\,,
\end{cases}
\ene
where $\gamma_{\psi}$ ($\gamma_{h}$) and $\beta_{\psi}$ ($\beta_{h}$) are the $\gamma$-factors and velocities of the top-quark pair system (Higgs boson) in the laboratory frame, respectively. Their ratio can be easily calculated,
\bee
\frac{ \gamma_{h}\beta_{h}  }{ \gamma_{\psi}\beta_{\psi} } 
=
\frac{ m_{\psi} }{ m_{h} } \gtrsim 2.77 \,,
\ene
where $m_{\psi}$ is the invariant mass of the top-quark pair, and the approximation stands for $m_{\psi} < 2m_{t}$ due to bound state effect near the threshold. Similarly, there is also an enhancement in the longitudinal component ($\lambda_{f}=0$), even through the scale factor is relatively small (the ratio is $\gamma_{h}/\gamma_{\psi} \approx 1.2$ at the threshold).

The enhancement effect does not only happen in the $CP$-odd observables, the $CP$-even observables are also strongly affected by the boost transformation. Fig.~\ref{fig:xs:evenThe}-\ref{fig:xs:evenThePhi} show the $\xi$ dependences of the above 9 $CP$-even asymmetry observables. The enhancement effect can be clearly seen in Fig.~\ref{fig:xs:evenThe}, \ref{fig:xs:evenTheThe} and \ref{fig:xs:evenThePhi} which correspond to distributions of the observables $\cala_{\theta_{i} }$, $\cala_{\theta_{1}\theta_{2}}$ and $\cala_{\theta_{i}\phi_{i'} }$, respectively.

Taking into account of the statistical uncertainty in the measurement of an asymmetry, the sensitivities of these asymmetries are estimated as follows,
\bee
\delta \cala_{i} \left(\delta \calc_{i}\right) = \sqrt{ \frac{ 1 - \cala_{i}^{2}(\calc_{i}^{2}) }{ \sigma_{SM} \epsilon \call } }
\ene 
where $\sigma_{SM}$ is the total cross section in the SM (as we have mentioned the number of events are always normalized to the SM prediction) and $\epsilon$ is the experimental efficiency factor, $\call$ is the integrated luminosity. At the production energy $\sqrt{s}=500{\rm GeV}$, the leading order (LO) cross section is calculated to be $\sigma_{\rm LO} = 0.29 {\rm fb}$ for the pure scalar case (we assume that the electron and position beams are not polarized). However, QCD correction is important in this energy region. After taking into account of the next LO and Coulomb resummation, the total cross section is about $\sigma_{\rm TOT} = 0.61 {\rm fb}$. Because the asymmetry observables are defined inclusively, the experimental efficiency can be assumed to be $1$ for a rough estimation. We also assume that the Higgs boson is identified by using the $h\to b\bar{b}$ channel which has a branching ratio about 56.9\%. The branching ratio of top quark to leptons $(\ell=e, \mu)$ is ${\rm Br}(t \to \ell X) = 19\%$. It is also important to note that while some asymmetries require identifications of both lepton and anti-lepton (for instance $\cala_{\theta_{1}\theta_{2}}$), there are also asymmetries that can be measured by using either one of them (for instance $\cala_{\theta_{1}}$). Therefore there are two kinds of configuration which we need to consider independently. For single lepton configuration, we have $\sigma_{SM}^{\ell} = 0.066{\rm fb}$. And it is $\sigma_{SM}^{\ell\ell} = 0.013{\rm fb}$ for double lepton configuration. For the projected luminosity $\call = 4 {\rm ab}^{-1}$ ~\cite{Barklow:2015tja}, we have $N_{SM}^{\ell} \approx 264$ and $N_{SM}^{\ell\ell} \approx 52$. Then the sensitivities of the single and double lepton asymmetries are roughly  
$\delta\cala^{\ell} \approx 0.06$ and $\delta\cala^{\ell\ell} \approx 0.14$, respectively. These sensitivities are not enough to have a decisive measurement on the $CP$ violating parameter. 
Further improvements can be obtained by 1) combining all these asymmetry observables; 2) slightly increasing the colliding energy, say $\sqrt{s}=550{\rm GeV}$~\cite{Hagiwara:2017ban}. 
Furthermore, since the asymmetries can be defined completely in the same way for the hadronic decay modes of the top-quark, the sensitivities can be enhanced significantly once the jets' charge can be distinguished~\cite{Barklow:2017suo} or more sophisticated method as shown in Ref.~\cite{Hagiwara:2017ban}. 

In summary, we have proposed a complete analysis method for probing possible $CP$-violation effects in the Yukawa interaction of top-quark at $e^{-}e^{+}$ collider. The observables are directly defined by using kinematical variables of the leptons from decays of the top-quarks, and hence reconstructions of the top-quarks are not necessary. Furthermore, these observables can be universally defined in any desired reference frame. Two reference frames, the rest frame of top-quark pair and the rest frame of Higgs boson, were examined. We find that observables that are trivial in the rest frame of the top-quark pair turn out to be sensitive to the $CP$-violating coupling in the rest frame of Higgs boson. Such enhancement effect happens for both $CP$-odd and $CP$-even observables, and particularly two new $CP$-odd observables were introduced based this effect. Our method can be directly used to improve the search for $CP$ violation effects at future $e^{-}e^{+}$ collider.

\section*{Acknowledgements}
K.M. thanks very much Prof. C.-P. Yuan for useful discussions. This study is supported by the National Natural Science Foundation of China under Grant No. 11705113, and Natural Science Basic Research Plan in Shaanxi Province of China under Grant No. 2018JQ1018, and the Scientific Research Program Funded by Shaanxi Provincial Education Department under Grant No. 18JK0153.

\bibliography{Phenomenology}

\end{document}